# Discontinuous epidemic transition due to limited testing

Davide Scarselli[1]†, Nazmi Burak Budanur[1]†, Marc Timme[2], Björn Hof[1]*

[1] Institute of Science and Technology Austria, Am Campus 1, 3400 Klosterneuburg, Austria

[2] Chair for Network Dynamics, Center for Advancing Electronics Dresden (cfaed), Technical University of Dresden, 01062 Dresden, Germany

*Correspondence to: bhof@ist.ac.at

†These authors contributed equally to this work

**High impact epidemics constitute one of the largest threats humanity is facing in the 21st century. Testing, contact tracing and quarantining are critical in slowing down epidemic dynamics, but may prove insufficient for highly contagious diseases. In the absence of pharmaceutical interventions, physical distancing measures remain as the last resort to avoid a widespread outbreak. Here we show that such combined countermeasures drastically change the rules of the epidemic transition if testing capacities are limited: Instead of continuous the response to countermeasures becomes discontinuous and rather than following the conventional exponential growth, the outbreak accelerates and scales super-exponentially during an intermediate period. As a consequence, containment measures either suffice to stop the outbreak at low total case numbers or fail catastrophically if marginally too weak, thus implying large uncertainties in reliably estimating overall epidemic dynamics, both during initial phases and during second wave scenarios.**

For high impact epidemics such as the ongoing COVID-19 pandemic, countries at least initially rely on non-pharmaceutical interventions to slow the outbreak dynamics. Keeping the maximum number of simultaneously infected individuals sufficiently low is of paramount importance to not overload health care system capacities[1,2]. Testing, quarantining and contact tracing have been combined with severe physical distancing measures across countries. Nevertheless, unlike, *e.g.*, for the 2002-2004 SARS outbreak[3] and the 2013-2016 Western African Ebola virus epidemic[4] such combined countermeasures could not yet stop the present COVID-19 pandemic. It thus remains an open question to date how testing, quarantining and contact tracing in combination with various physical distancing measures affect the epidemic dynamics, and in particular the epidemic peak that represents a worst case scenario regarding the pressure on the health care system.

Researchers and policy makers often implicitly assume that the peak, i.e. the largest fraction of simultaneously infected individuals, continuously varies with epidemic parameters and with the level of countermeasures implemented. In this article, we demonstrate that this fundamental assumption is incorrect once testing resources are limited. We reveal that the nature of the epidemic dynamics changes drastically from this naive picture and has unexpected, severe consequences. In particular, limited testing generically yields a discontinuous transition in the fraction of infected individuals in a population, a phenomenon dynamically accompanied by an interval of super-exponential growth. Similar to related types of phase transitions in statistical physics such as discontinuous or explosive percolation transitions[5–8], limited testing effectively delays the transition, such that the fraction of infected individuals explosively becomes macroscopically large once effective epidemic parameters even only marginally cross a threshold. As a consequence, in the presence of limited testing, slight changes in countermeasures may induce huge macroscopic changes in the fraction of infected individuals, thereby making the transition highly unpredictable.

In many epidemic models, such as the SIR model, the population is commonly considered large and divided into compartments such as susceptible (*S*), infectious (*I*) or recovered (*R*) and the evolution of these compartments is traditionally[9] modeled by ordinary differential equations (ODEs). For small populations, number fluctuations become relevant such that stochastic, microscopic network models are more appropriate, whereas for increasingly large populations, deterministic mean field approaches are usually considered suitable because relative fluctuations in the susceptible, infectious etc. populations become less and less important. In comparison to more complex models that take into account population structure and stochasticity, ODE models are often also motivated by the simplicity of implementation as well as by the greater ease in analyzing and interpreting the results. With this perspective in mind, in large-scale outbreaks such as the ongoing COVID-19 pandemic, ODE models would be expected to capture the overall general features of the epidemic dynamics.

However, as we explain below, if a disease spreads despite intervention, large populations size and large numbers of susceptible and infectious individuals and large numbers of tests etc. still constitute insufficient conditions for neglecting fluctuations, because the key quantity during the early growth phase of an epidemic is the difference between two of these numbers, the number $N_T$ of available tests each day and the number $N_S$ of individuals suspected to carry the disease and thus (ideally) to be tested. Regardless of the overall population size, the total number of susceptible and other macroscopic population numbers, the difference $\Delta_{\text{Test}}=N_T - N_S$ may be or become small and thus introduce relevant fluctuations, and if it becomes negative the epidemic spread subsequently accelerates, *i.e.* during this phase the growth is faster than exponential. Precisely this effect alters the nature of the epidemic transition and makes it discontinuous. A small variation of the epidemic parameters does not cause the expected small change to the growth process but yields disproportionate consequences with an explosive increase of the fraction of infected individuals in the population.

Let us consider a basic dynamic agent-based model that simultaneously captures the epidemic dynamics and the influence of countermeasures and has previously been studied for analyzing the dynamics of Ebola[10]. For simplicity, we initially focus on two dimensional square lattice grids where each agent or tile represents an individual and interactions can be two-fold, either short-range via nearest-neighbor contacts or long range representing out-of-neighborhood contacts, for instance mediated by agent mobility. A tile falls into one of four compartments, susceptible (*S*), exposed (*E*), infectious (*I*) or recovered (*R*) resulting in an *SEIR* model. In addition, we split the infectious population into two categories, strong- and weak-symptom cases, where the latter represent individuals with either unspecific or no symptoms. As a result of intervention measures, each of above states can be put under quarantine, formally increasing the number of possible states to eight (for a detailed description of all the states and possible interactions see Supplementary Fig. 1).

Key features of the model can be understood from the example illustrated in Fig. 1. We consider discrete time dynamics with time steps representing days. Strong-symptom individuals (red) are immediately identified and automatically quarantined (blue dashed border). After testing positive, its four nearest neighbors are quarantined and queued for testing. For every new positive case, the quarantining and testing procedure is continued. In this simple scenario, all local contacts (four neighbors) are traced, however, prior interactions with distant sites are assumed untraceable. Weak-symptom cases (brown tile bottom right of Fig.1) go undetected unless identified through contact tracing. At each time step they can spread the disease with a constant probability to the four nearest neighbors plus to a randomly chosen distant site (mimicking random encounters, *e.g.* during travel).

We start the simulations from a small number of weak-symptom infectious randomly scattered across a population of $P=3162\times3162\approx10^7$. We model the incubation and infectious period with a Gamma distribution with parameters are similar to the ones reported[11] for COVID-19. The transmission probability is set to 0.38 to reproduce the average growth rate $\sim\exp(\kappa t)$ with $\kappa = 0.3$ day$^{-1}$ observed during the early exponential phase of the ongoing COVID-19 pandemic. In addition, we assume 50% of the new infections to show only weak symptoms. While the exact ratio of weak-symptom carriers of COVID-19 is unknown, their prevalence is reported in multiple studies[12–15]. Moreover, to allow for a realistic testing scenario we set an upper limit of daily tests of $10^{-4}$ $P$ (*i.e.* 1000 tested individuals per day). This limit was chosen since it approximately corresponds to the largest fraction of the population tested in any European country during the COVID-19 outbreak[16]. In most countries, the daily tests conducted were significantly lower during the early phase of the epidemic. All results reported below are robust against the specific values of $N_T$ as long as the daily test limit is significantly smaller than the total population size ($N_T<<P$). This parameter choice results in a basic reproduction number, $R_0\approx3$, and an outbreak (leftmost curve in Fig. 2a) with such a high transmission rate can in general (*i.e.* in the thermodynamics limit) not be halted by above testing and contact tracing intervention scheme.

We next consider how the outcome of the epidemic is altered if the testing and contact tracing interventions are aided by additional mitigation measures (*e.g.* social distancing). Unlike testing and quarantining which are simulated directly as described above, additional mitigation measures are modeled by a reduction in the transmission rate. In order to investigate the response to different levels of mitigation the transmission rate is reduced, which translates to a continuous decrease in the basic reproduction number. As shown in Fig. 2a the epidemic curve flattens at first continuously as the mitigation strength increases. However once the basic reproduction number marginally drops below a value of 2.5 the epidemic peak discontinuously drops to a very low value. Hence for $R_0<2.5$ the outbreak is halted (the fraction of infected tends to zero in the thermodynamic limit). In contrast the familiar continuous picture of flattening the curve is recovered when testing interventions are removed as shown in Fig. 2b. Here a continuous reduction in $R_0$ causes the expected continuous decrease of the peak of the epidemic curve (see also Supplementary Video 1).

The discontinuity in the presence of testing and contact tracing is equally apparent when considering the total number of infected at the end of the epidemic ($N_F$), shown for decreasing mitigation strength (*i.e.* increasing $R_0$) in Fig. 2c. While testing and contact tracing can suppress outbreaks with basic reproduction numbers significantly larger than one, once containment fails it does so catastrophically, *i.e.* the fraction of the population eventually infected jumps from close to zero directly to a large fraction, in this case approximately $0.5P$. The cause of the discontinuous response can be understood from the time evolution of the empirically-computed[10] effective reproduction number $R_t$. As shown in Fig. 2d, for a suppressed outbreak ($R_0=2.3$, black circles) testing and contact tracing reduce the reproduction number to just below one and hence the number of infectious decreases exponentially and the outbreak is eventually suppressed. For $R_0$=2.7 however the effective reproduction number can only be reduced to a value slightly larger than one. Consequently the number of infectious increases exponentially.

So far the difference between these two cases is exactly as standard models would predict. As time proceeds however, in the latter case the number of suspects will eventually reach the test capacity limit, $\Delta_{\text{Test}}$=0 (at this point the positive rate increases see Supplementary Fig. 2). Subsequently, a fraction of the infectious are only tested with a delay and therefore have a larger probability to transmit the disease. As shown in Fig. 2d this leads to an increase in the reproduction number and hence the outbreak accelerates. Once set into motion the number of unchecked suspects continues to increase and so does the reproduction number. Instead of the familiar exponential growth during epidemics, the growth at this stage is super-exponential (see Supplementary Fig. 3) because $R_t$ and thereby the exponent of the growth dynamics, increases with time. A marginal difference in $R_0$ (compared to the suppressed case) is amplified into a significant difference in the effective reproduction number $R_t$ once the contact tracing capacity limit is exceeded. It is precisely this basic amplification mechanism that turns flattening the epidemic curve into a process with discontinuous overall outcome that arises explosively, *i.e.* without prior warning signs. This mechanism is not specific to the simple model chosen here but is equally found in small world[17,18] or scale free networks[19] (see Supplementary Fig. 4) and it is independent of details in the epidemic dynamics. The mechanism and the induced discontinuous transition thus generally emerges if testing and contact tracing have an upper capacity limit. The direct observation of such acceleration is difficult to observe in practice, because testing is overwhelmed and does not reflect the actual numbers. For the same reason the time dependent reproduction number $R_t$ is strongly fluctuating and its true value is hard to estimate from observed time series.

A limited testing capacity does not only alter the response to mitigation during the early stages of an epidemic but equally introduces a discontinuity when considering lock down scenarios of varying strength. For a simple illustration we again simulate an outbreak in a population of size $P$=3162×3162≈$10^7$ that has spread to $10^4$ infectious. At this point we assume a basic reproduction number of $R_0$=2. This could be either interpreted as a less contagious disease, or as the same disease as in the previous case, where the reproduction number has been reduced by social distancing measures. At this stage of the outbreak the number of suspects in the population already far exceeds the number of daily available tests ($\Delta_{\text{Test}}$<0) and contact tracing cannot suppress the outbreak. To get the situation back under control strong mitigation measures (*i.e.* a lock down) are required and we consider that as a result the basic reproduction number is further decreased by 0<$\Delta R_0$<2. In each case the lock down is enforced for 30 days. As shown in Fig. 3, for sufficiently strong lock downs ($\Delta R_0$> $\Delta R_c$≈0.7) the outbreak is eventually suppressed (*i.e.* the effective reproduction number is reduced below one). However, if the lock down is just marginally weaker and $\Delta R_0$< $\Delta R_c$ containment catastrophically fails: At the end of the lock down the number of suspects still (marginally) exceeds the test capacity and as time proceeds $R_t$ increases (not shown). While in Fig. 2a the discontinuity separates epidemics subject to different mitigation levels, here the discontinuity arises from the difference in the number of active cases at the end of the lock down.

In common epidemic models testing and contact tracing are often incorporated in the basic reproduction number, yet that approach does not take into account capacity limits and hence cannot reflect scenarios in which such capacities are eventually exhausted as the epidemic continues to spread. As we have demonstrated, however, including such capacity limits drastically alters the overall epidemic dynamics and thus need to be carefully considered both in research and for policy making. This holds in particular during early growth periods of the outbreak as well as during potential second waves.

While during the COVID-19 pandemic the focus has been on the key role of mitigation in protecting the health care systems, the above results indicate that additional mitigation measures may play an equally vital role in protecting the efficiency of testing and contact tracing. If it fails, no matter how marginally, the disease begins to spread at an accelerating rate and will in due course infect a large fraction of the population. In practice, the eventual outcome of the epidemic might still be averted, if countermeasures are severely strengthened quickly after this acceleration. However, the suppression of the outbreak comes at a significantly higher cost, since more stringent mitigation measures are required to regain control.

**References and Notes:**

1. Lilienfeld, D. E. & Stolley, P. D. *Foundations of Epidemiology*. (Oxford University Press, USA, 1994).
2. Sands, P., Mundaca-Shah, C. & Dzau, V. J. The neglected dimension of global security—a framework for countering infectious-disease crises. *N. Engl. J. Med.* **374**, 1281–1287 (2016).
3. Riley, S. & Al., E. Transmission dynamics of the etiological agent of SARS in Hong Kong: impact of public health interventions. *Science (80-. ).* **300**, 1961–1966 (2003).
4. Saurabh, S. & Prateek, S. Role of contact tracing in containing the 2014 Ebola outbreak: a review. *Afr. Health Sci.* **17**, 225–236 (2017).
5. Achlioptas, D., D'Souza, R. M. & Spencer, J. Explosive percolation in random networks. *Science (80-. ).* **323**, 1453–1455 (2009).
6. Nagler, J., Levina, A. & Timme, M. Impact of single links in competitive percolation. *Nat. Phys.* **7**, 265–270 (2011).
7. D'Souza, R. M. & Nagler, J. Anomalous critical and supercritical phenomena in explosive percolation. *Nat. Phys.* **11**, 531–538 (2015).
8. Fan, J. *et al.* Universal gap scaling in percolation. *Nat. Phys.* **16**, 455–461 (2020).
9. Kermack, W. O., McKendrick, A. G. & Walker, G. T. A contribution to the mathematical theory of epidemics. *Proc. R. Soc. London. Ser. A, Contain. Pap. a Math. Phys. Character.* **115**, 700–721 (1927).
10. Wong, V., Cooney, D. & Bar-Yam, Y. Beyond Contact Tracing: Community-Based Early Detection for Ebola Response. *PLOS Curr. Outbreaks* **1**, (2016).
11. Bar-On, Y. M., Flamholz, A., Phillips, R. & Milo, R. Science Forum: SARS-CoV-2 (COVID-19) by the Numbers. *Elife* **9**, e57309 (2020).
12. Gandhi, M., Yokoe, D. S. & Havlir, D. V. Asymptomatic Transmission, the Achilles' Heel of Current Strategies to Control Covid-19. *N. Engl. J. Med.* **382**, 2158–2160 (2020).
13. Furukawa, N. W. & Brooks, J. T. Evidence Supporting Transmission of Severe Acute Respiratory Syndrome Coronavirus 2 While Presymptomatic or Asymptomatic. *J. Sobel, . Emerg. Infect. Dis.* **26**, (2020).
14. Mizumoto, K., Kagaya, K., Zarebski, A. & Chowell, G. Estimating the Asymptomatic Proportion of Coronavirus Disease 2019 (COVID-19) Cases on Board the Diamond Princess Cruise Ship, Yokohama, Japan, 2020. *Euro Surveill.* **25**, (2020).
15. Day, M. Covid-19: Four Fifths of Cases Are Asymptomatic, China Figures Indicate.

*BMJ* **369**, m1375 (2020).

16. Roser, M., Ritchie, H., Ortiz-Ospina, E. & Hasell, J. Coronavirus Pandemic (COVID-19). *Publ. online OurWorldInData.org* (2020).

17. Watts, D. J. & Strogatz, S. H. Collective dynamics of 'small-world' networks. *Nature* **393**, 440–442 (1998).

18. Kleinberg, J. M. Navigation in a Small World. *Nature* **406**, 845 (2000).

19. Barabási, A.-L. & Albert, R. Emergence of scaling in random networks. *Science (80-. ).* **286**, 509–512 (1999).

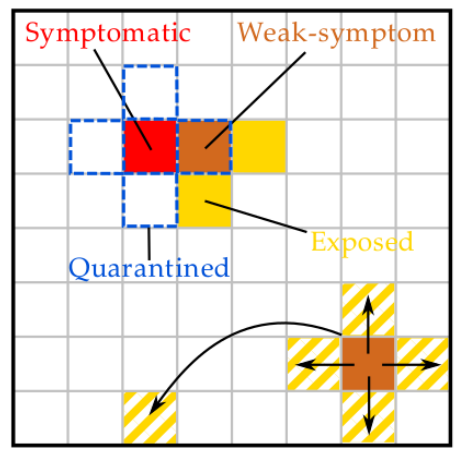

**Fig. 1 Spatial epidemic model with testing and quarantining.** Every day each infectious individual (agent represented here as a tile on a lattice) interacts with their neighbors and a randomly selected individual, and transmits the disease (arrows in figure) with constant probability if the individuals they interact with are susceptible. The tiles with yellow and white stripes denote the potential contacts that can be exposed. Upon identification of a positive case (red tile) all the neighbors are put into quarantine and tested (blue dashed border). Weak-symptom cases (brown tile with blue dashed border) can only be identified if they are neighbors of a known positive case.

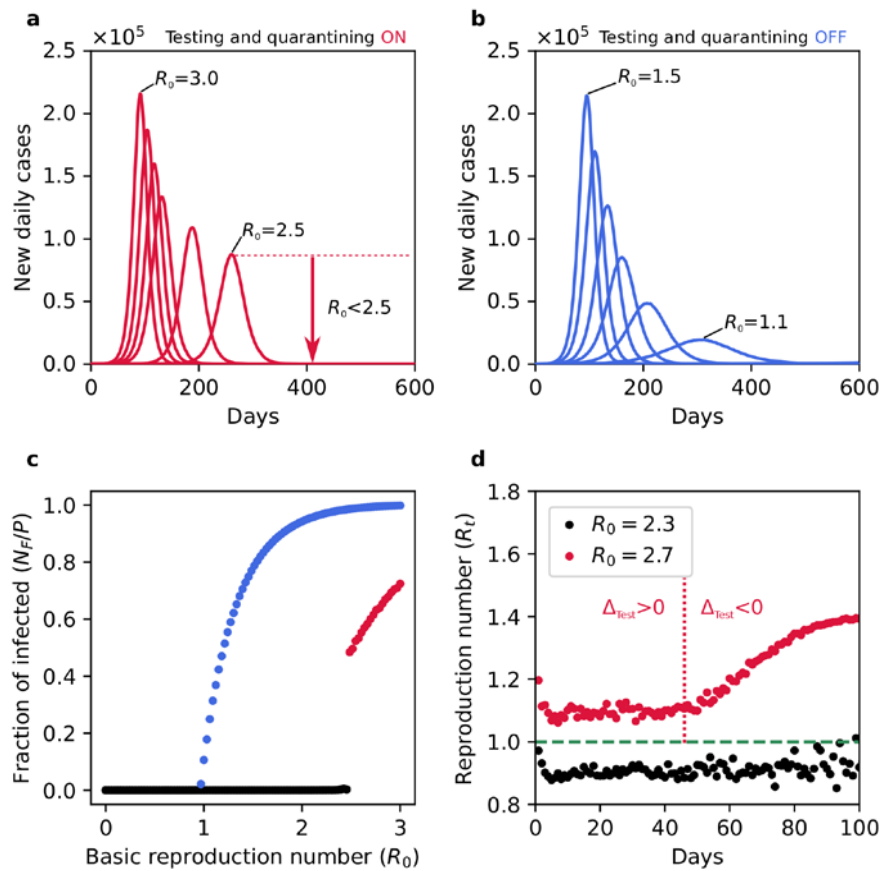

**Fig. 2 Discontinuity in flattening of epidemic curves. a** Daily new cases for continuously decreasing values of $R_0$ ($3 > R_0 > 0$), mimicking mitigation measures of increasing strength. Testing and quarantining are carried out at the same time with a capacity limit of $N_T = 1000$ tested individuals per day. Initially the peak reduces continuously in response to mitigation (red curves from left to right), however once $R_0$ is reduced below 2.5 the epidemic curve drops to very small numbers of new cases (curves not visible in the figure scale). **b** Daily new cases for decreasing values of $R_0$ ($1.5 > R_0 > 0$) without testing and quarantining. Decreasing progressively $R_0$ gradually flattens the epidemic curve (blue curves from left to right) until a very low number of cases is reached (curves not visible in the figure scale). **c** Final fraction of infected ($N_F/P$) as a function of $R_0$ corresponding to the curves shown in **a** (red dots) and **b** (blue dots). The black dots denote outbreaks that have been effectively suppressed. When testing and quarantining are active the epidemic transition becomes discontinuous at happens

at approximately at a higher basic reproduction number (approximately $R_0$=2.5) with respect to the usual continuous epidemic transition observed at $R_0 = 1$. **d** Evolution of the reproduction number with testing and quarantining active. This intervention can efficiently reduce the reproduction number $R_t$ below one for $R_0 < 2.3$ (black dots). For larger values of $R_0$ testing and quarantining can initially reduce the reproduction number to a constant level, however $R_t$ remains above one (red dots). Due to the continuing spread the number of suspects will eventually exceed the daily test limit and hence $\Delta_{Test}$ changes sign (red dashed line in **d**). At this point the spread accelerates and $R_0$ increases. The values of $R_t$ have been averaged over 800 simulations for each case ($R_0$=2.3 and $R_0$=2.7). In all these cases the population $P$=3162×3162≈$10^7$ people and epidemics start with 100 initial infectious.

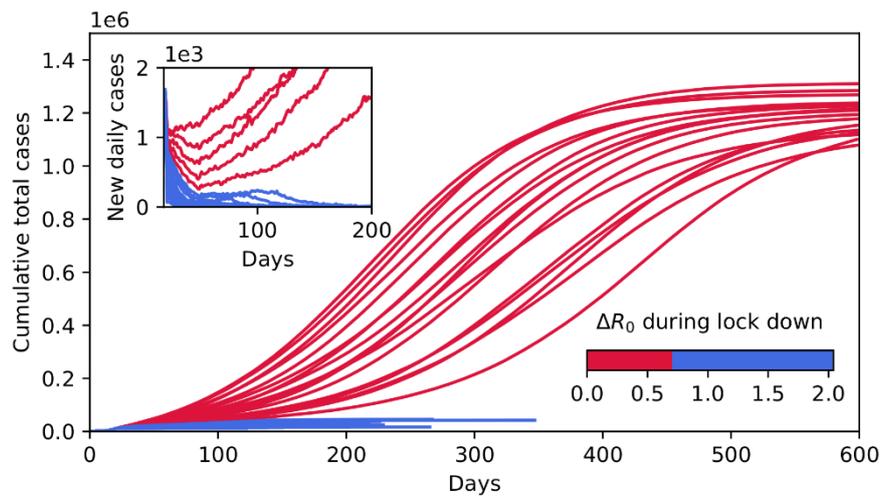

**Fig. 3 Discontinuous lock down scenarios.** Cumulative total cases for $R_0 = 2$ starting from $10^4$ infectious in a population $P$=3162×3162≈$10^7$ and subjected to gradually stronger lock downs. Mitigation measures are simulated by a reduction in the basic reproduction number in the range of 0<$\Delta R_0$<2.0. The duration is 30 days in all cases. Testing is limited to $N_T$=1000 individuals per day. Mild interventions (0<$\Delta R_0$<0.7, red curves) only result in an initial drop (see inset) in daily new cases but ultimately cannot prevent a subsequent rise in numbers and eventually a high proportion of the population becomes infected. Stronger interventions (0.7<$\Delta R_0$<2.0, blue curves) on the other hand efficiently bring the epidemic under control. Inset, corresponding epidemic curves of daily new cases. Reducing continuously $R_0$ during lock down produces a family of epidemic curves that ultimately result in a discontinuous outcome: Either the outbreak is suppressed (blue curves), or containment fails catastrophically leading to a high proportion of the population being infected (red curves).

**Methods:**

Spatial epidemic model

Our base model is a spatial *SEIR* (Susceptible-Exposed-Infectious-Recovered) model, in which a population $P=3162\times3162$ is represented by a 2-dimensional grid where each grid point represents an individual. In addition to the above four compartments we distinguish between symptomatic ($I_S$) and weak-symptom ($I_W$) individuals, where the latter ranges from people who may have unspecific symptoms (e.g. coughing) to entirely asymptomatic. With the introduction of intervention measures aimed at containing the disease spread, the individuals in the states $S$, $E$, $I_S$ and $I_W$ can be put under quarantine ($Q_S$ for $S$, $Q_E$ for $E$ and $Q_I$ for $I_S$ and $I_W$). All the eight states and the possible transition paths are shown in Supplementary Fig. 1 and described in the caption. Simulations start from a small group of 100 IW that are randomly scattered across the grid. Each infectious ($I_S$ or $I_W$) is assigned an infectious period which is drawn from a Gamma distribution with mean 4 days and one day as scale parameter. During the infectious period these individuals can interact with each of the four neighbors and a randomly chosen additional individual. The disease is transmitted with a given probability, if the target individuals are susceptible (cf. Fig. 1). After the infectious period $I_S$ and $I_W$ transform into recovered ($R$) and can not interact any more with the population. Once a susceptible individual is infected, it transforms into exposed ($E$) and is assigned an incubation period which is drawn from a Gamma distribution with mean 3 days and one day as scale parameter. After the incubation period is elapsed the state of the individual is transformed from $E$ into $I_W$ with probability $p_W = 0.5$. For a large population size $p_W$ thus represents the ratio of weak-symptom cases to the infected population. The transmission probability is chosen to reproduce the average growth rate observed during the early exponential phase of the ongoing COVID-19 pandemic. To this end, we run several simulations without any containment and we pick the transmission probability that minimizes the difference from the growth rate $\sim\exp(\kappa t)$ with $\kappa = 0.3$ day$^{-1}$.

Testing and quarantining model

The implementation of the epidemic mitigation is based on identification, quarantining and testing of suspect cases (cf. Fig. 1 for an illustrative cartoon of the process). The response starts when a first symptomatic case ($I_S$) appears and is recognized as suspect case. The individual is immediately quarantined and tested. Upon the positive test result, the status is switched to $R$ and its neighbors are quarantined and queued for testing. Each day, $N_T$ (daily available number of tests) individuals in the queue are tested. The test outcome is revealed with a delay of one day and the same known positive cases can not be used more than once for tracing its neighbors. In case of negative test ($Q_S$) the individual is reverted to susceptible.

Alternative networks

The model can be easily extended to a different network structure while retaining the state transition rules and parameters. In order to assess the robustness of our results we considered two additional networks (the results are shown in Supplementary Fig. 3). For the first one we chose Kleinberg's Navigable Small World[9,10] as implemented in NetworkX 2.4 Python library. Here each individual is connected to a random person on the grid, with the probability of being connected to a person decreasing as $\sim d^{-2}$, where $d$ is the taxicab distance over the grid. Moreover, the connection is static and it is not assigned on a daily basis. In the second model we adopt a fully scale-free network[11] with the number of connections per person drawn from a discrete zeta distribution with parameter 2 and cutoff 100, and all are static and do not change during the simulation.

**Author Contributions:**

D.S., N.B.B. and B.H. designed the epidemic model with testing and quarantining. D.S. and N.B.B. wrote the Python code and performed the numerical simulations. D.S., N.B.B. M.T. and B.H. analysed, interpreted the results and wrote the paper.

**Competing Interests:**

The authors declare no competing financial interests.